\documentclass[amsmath,amssymb,aps,pra,reprint,superscriptaddress]{revtex4-2}

\usepackage{graphicx}
\usepackage{dcolumn}
\usepackage{bm}
\usepackage{multirow}
\usepackage{url}

\usepackage{pdfcomment}
\usepackage{xcolor}
\usepackage{algorithm}
\usepackage{algpseudocodex}
\usepackage{caption}

\usepackage{subfigure}
\begin{document}

\title{Iterative Partition Search Variational Quantum Algorithm for Solving Shortest Vector Problem}

\affiliation{State Key Laboratory of Networking and Switching Technology, Beijing University of Posts and Telecommunications, Beijing 100876, China}
\affiliation{State Key Laboratory of Cryptology, P. O. Box 5159, Beijing 100878, China}
\affiliation{National Key Laboratory of Security Communication, Institute of Southwestern Communication, Chengdu 610041, China}
\affiliation{School of Cyberspace Security, Beijing University of Posts and Telecommunications, Beijing 100876, China}

\author{Zi-Wen Huang}

\affiliation{State Key Laboratory of Networking and Switching Technology, Beijing University of Posts and Telecommunications, Beijing 100876, China}
\affiliation{State Key Laboratory of Cryptology, P. O. Box 5159, Beijing 100878, China}
\affiliation{School of Cyberspace Security, Beijing University of Posts and Telecommunications, Beijing 100876, China}

\author{Xiao-Hui Ni}
\affiliation{State Key Laboratory of Networking and Switching Technology, Beijing University of Posts and Telecommunications, Beijing 100876, China}
\affiliation{School of Cyberspace Security, Beijing University of Posts and Telecommunications, Beijing 100876, China}
\author{Jia-Cheng Fan}
\affiliation{State Key Laboratory of Networking and Switching Technology, Beijing University of Posts and Telecommunications, Beijing 100876, China}
\affiliation{School of Cyberspace Security, Beijing University of Posts and Telecommunications, Beijing 100876, China}
\author{Su-Juan Qin}
\affiliation{State Key Laboratory of Networking and Switching Technology, Beijing University of Posts and Telecommunications, Beijing 100876, China}
\affiliation{School of Cyberspace Security, Beijing University of Posts and Telecommunications, Beijing 100876, China}
\author{Wei Huang}
\email{huangwei096505@aliyun.com}
\affiliation{National Key Laboratory of Security Communication, Institute of Southwestern Communication, Chengdu 610041, China}
\author{Bing-Jie Xu}
\affiliation{National Key Laboratory of Security Communication, Institute of Southwestern Communication, Chengdu 610041, China}
\author{Fei Gao}
\email{gaof@bupt.edu.cn}
\affiliation{State Key Laboratory of Networking and Switching Technology, Beijing University of Posts and Telecommunications, Beijing 100876, China}
\affiliation{State Key Laboratory of Cryptology, P. O. Box 5159, Beijing 100878, China}
\affiliation{School of Cyberspace Security, Beijing University of Posts and Telecommunications, Beijing 100876, China}

\date{\today}

\begin{abstract}
The Partition Search Algorithm (PSA) and Iterative Quantum Optimization with an Adaptive Problem (IQOAP) are leading variational quantum algorithms for solving Shortest Vector Problem (SVP).
However, each has limitations that restrict its practical impact.
IQOAP suffers from ineffective iterations that fail to update the lattice basis, whereas PSA's static partitioning leads to oversized search spaces.
In this work, we propose the Iterative Partition Search Algorithm (IPSA), which systematically addresses these drawbacks by integrating a ``1-tailed search spaces" with a dynamic, stack-managed iterative process.
Specifically, the ``1-tailed" strategy ensures that every successful execution yields an effective lattice basis update, thereby eliminating the ineffective iterations associated with IQOAP.
Concurrently, the dynamic iterative process reduces the required qubit count, thereby avoiding the limitation of an oversized search space inherent to PSA.
We validate IPSA on the \textit{Baihua} superconducting quantum processor via the Quafu platform.
Small-scale real hardware experiments demonstrate that, compared to PSA, IPSA achieves a 14-fold increase in success rate at a cost of less than double the total circuit depth.
Conversely, compared to IQOAP, IPSA reduces the total circuit depth by 82.7\% while achieving approximately 2.5 times its success rate.
Furthermore, we also conduct numerical simulations whose results are in good agreement with the experimental findings and extend our analysis.
\end{abstract}
\maketitle

\section{~\label{sec:intro}Introduction}
The Shortest Vector Problem (SVP) is an NP-hard problem in lattice theory, underpinning the security of numerous post-quantum cryptographic schemes.
Lattice is defined as a discrete additive subgroup of ${\mathbb{R}}^{n}$, typically generated by a basis matrix.
These basis vectors form a lattice through their integer linear combinations.
SVP seeks to identify the shortest non-zero vector within a lattice structure.
Classical algorithms such as enumeration~\cite{kannan1983improved,fincke1985improved,gama2010lattice} and sieving~\cite{ajtai2001sieve,micciancio2010faster,becker2016new} can solve SVP but require substantial computational costs.
Quantum algorithms offer promising alternatives by exploiting Grover's search~\cite{grover1996fast} and quantum walks~\cite{laarhoven2015finding,chailloux2021lattice,bonnetain2023finding,aono2018quantum} to achieve significant speedups.
Recent quantum advances have reduced sieving complexity to $2^{0.2563n+o(n)}$~\cite{bonnetain2023finding} and demonstrated quadratic speedups for enumeration method~\cite{aono2018quantum}.

However, these quantum advantages typically require fault-tolerant quantum computers, which remain beyond current technological capabilities.
Present-day Noisy Intermediate-Scale Quantum (NISQ) devices are characterized by limited qubit counts and inherent noise, motivating the development of Variational Quantum Algorithms (VQAs)~\cite{cerezo2021variational,lubasch2020variational,liu2021variational,wang2021variational,bravo2023variational}.
VQA represents hybrid quantum-classical computing paradigms that employ parameterized quantum circuits, described by parameters $\bm{\theta}$, to prepare trial states $|\psi(\bm{\theta})\rangle$.
A classical optimizer iteratively adjusts $\bm{\theta}$ to minimize a cost function $C(\bm{\theta})=\langle\psi(\bm{\theta})|H|\psi(\bm{\theta})\rangle$, where $H$ represents a Hamiltonian encoding the optimization problem.
With a shallower circuit and inherent noise tolerance~\cite{mcclean2016theory}, VQA is well-suited for NISQ devices, with prominent examples ranging from foundational algorithms like the Variational Quantum Eigensolver (VQE)~\cite{peruzzo2014variational} and the Quantum Approximate Optimization Algorithm (QAOA)~\cite{farhi2014quantum}, to broader applications in areas like quantum federated learning~\cite{song2024quantum} and quantum neural networks~\cite{wu2024quantum,li2025efficient,su2025topology}.

Investigating VQA-based solvers for SVP is a meaningful endeavor, as it helps assess the potential threat that NISQ devices pose to post-quantum cryptography.
Applying VQA to SVP involves 
two
key aspects of consideration, and several existing studies have begun to address each of these points.

First, mapping the infinite coefficients search space $\mathbb{Z}^n$ of SVP onto finite qubit count requires binary encoding strategy~\cite{joseph2021two} and the establishment of bounded search ranges~\cite{albrecht2023variational,zheng2025quantum}, where the range boundaries are typically guided by classical preprocessing algorithms such as Lenstra–Lenstra–Lovász (LLL)~\cite{lenstra1982factoring} and Hermite-Korkin-Zolotarev (HKZ)~\cite{kannan1983improved}.
Recently, Zhu, Joseph et al. proposed the Iterative Quantum Optimization with an Adaptive Problem (IQOAP) framework for solving the SVP~\cite{zhu2022iterative}, achieving qubit requirement reduction compared with conventional boundary constraint methods~\cite{albrecht2023variational,zheng2025quantum}.

Second, the straightforward Hamiltonian construction for the vector norm objective leads to a trivial all-zero ground state, necessitating sophisticated approaches to avoid this trivial solution.
Standard avoidance strategies, such as introducing penalty terms and redesigning the cost function, come with certain costs due to the difficulty in tuning penalty coefficients and the requirement for additional qubits~\cite{albrecht2023variational,mizuno2024quantum,barbera2024finding}.
The Partition Search Algorithm (PSA)~\cite{yamaguchi2022annealing}, originally proposed and validated in adiabatic quantum computing, offers an elegant approach to bypass this trivial solution issue without incurring any additional costs.
Although initially developed within the adiabatic quantum computing framework, PSA demonstrates equivalent efficacy in VQA for avoiding trivial solutions.

However, despite their respective contributions, both PSA and IQOAP possess inherent limitations that hinder their practical performance.
For PSA, its static partitioning strategy requires a high number of qubits, which in turn generates an oversized search space. This vast search space frequently traps the VQA in suboptimal local minima due to the complex optimization landscape.
IQOAP, on the other hand, does not incorporate any mechanism to circumvent the trivial all-zero solution. Furthermore, its iterative structure is prone to ineffective iterations, in which a successfully found shorter vector is discarded because it fails to meet the requirements for a lattice basis update, thereby wasting computational resources.

To address these issues, we propose the Iterative Partition Search Algorithm (IPSA), which systematically addresses these drawbacks by integrating ``1-tailed search spaces" into a dynamic, stack-managed iterative process.
The dynamic iterative process directly addresses the limitations of PSA's static approach. By enabling the use of smaller, successive search spaces, it fundamentally avoids the vast and complex optimization landscape that plagues PSA.
Concurrently, the ``1-tailed search spaces" resolve the primary drawbacks of IQOAP. It inherently avoids the trivial all-zero solution and ensures that every successfully found shorter vector meets the requirements for a lattice basis update, thereby eliminating ineffective iterations.
As a result, IPSA demonstrates a superior trade-off between solution quality and computational cost compared to its predecessors, further advancing the application of VQA for solving the SVP.

The remainder of this paper is organized as follows. In Section~\ref{sec:methodAB_merged}, we briefly review two existing methods, the PSA and the IQOAP. In Section~\ref{sec:methodD}, we present our proposed IPSA and detail its key components, including the 1-tailed search spaces and the stack-managed iterative process of search partitions.
Section~\ref{sec:hardware_exp} reports the results of our experiments on real quantum hardware.
In Section~\ref{sec:con}, we provide our conclusions and a brief outlook for future applications.
A discussion justifying the selection of our parameterized quantum circuit is presented in Appendix~\ref{app:A}. Detailed information on the quantum processor's parameters and topology is provided in Appendix~\ref{app:B}. Further numerical simulations are presented in Appendix~\ref{app:C}.

\section{~\label{sec:methodAB_merged}Foundation: Partition Search and Iterative Optimization Framework}

The PSA~\cite{yamaguchi2022annealing} addresses the zero-vector issue without incurring additional computational overhead. It partitions the coefficient space of basis vector combinations into non-overlapping regions and solves the SVP independently within each region. For an $n$-dimensional lattice, the coefficient space is divided into regions $X_1, \ldots, X_n$, where each region $X_i$ is defined as:
\begin{equation}
~\label{eq:xi}
X_i = \{(x_1,\dots,x_i,0,\dots,0)^T \in\mathbb{Z}^n : x_i \ge 1\}
\end{equation}
This partitioning ensures that regions are mutually exclusive and none contains the zero vector, thereby eliminating the trivial solution.

To construct the problem Hamiltonian for each region $X_i$, we assume each coefficient is represented using binary encoding with appropriate qubit allocation. The first $i-1$ coefficients are encoded as:
\begin{equation}
\hat{x_r}=\frac{1}{2}-\sum_{j=0}^{m}2^{j-1}\sigma^z_{r,j}
\end{equation}
representing integers in the range $(-2^m, 2^m]$, where $\sigma^z_{r,j}$ denotes the Pauli-Z operator acting on the $j$-th qubit of the $r$-th coefficient. The $i$-th coefficient, constrained by $x_i \ge 1$, is encoded as:
\begin{equation}
\hat{x_i}=\sum_{j=0}^{m-1}2^{j-1}(\sigma^z_{i,j}+1)+1
\end{equation}
representing integers in the range $[1, 2^m]$. The Hamiltonian for region $X_i$ is constructed as the squared Euclidean norm:
\begin{equation}
~\label{eq:Hi}
H_i=\sum_{k=1}^{n}\left(\sum_{r=1}^{i-1}\hat{x_r}b_{r,k}+\hat{x_i}b_{i,k}\right)^2
\end{equation}
where $b_{r,k}$ denotes the $k$-th component of the $r$-th lattice basis vector.

The ground state of each $H_i$ yields the shortest vector within the corresponding partition, and the global minimum among all partitions provides the SVP solution. The total qubit requirement $M$ depends on the sum of bits needed to represent each coefficient $m$, which is determined by the preprocessing of the lattice basis. The analysis in Ref.~\cite{albrecht2023variational} shows that for an HKZ-reduced basis, the requirement is $M = \mathcal{O}(n \log n)$, while Ref.~\cite{zheng2025quantum} concludes that using an LLL-reduced basis requires $M = n(n+1)$ qubits.

To reduce the qubit utilization in boundary constraint methods, the IQOAP framework~\cite{zhu2022iterative} performs VQA optimization within a smaller, fixed coefficient range, leveraging the fact that even suboptimal solutions can progressively reduce the lattice basis through iterative refinement. For example, in 4-dimensional SVP instances, each $x_r$ is encoded using only $2$ qubits via an encoding like:
\begin{equation}
~\label{eq:xrprime}
\hat{x_r^\prime}=\frac{1-\sigma^z_{r,0}-2\sigma^z_{r,1}}{2}
\end{equation}
Correspondingly, the Hamiltonian expression is:
\begin{eqnarray}
~\label{eq:Hprime}
H^\prime=\sum_{k=1}^{n}(\sum_{r=1}^{n}\hat{x_r^\prime}b_{r,k})^2
\end{eqnarray}

The IQOAP framework improves the basis quality through an iterative refinement process, which is outlined in Algorithm~\ref{alg:IQOAP}. In each iteration, if the VQA finds a vector $\bm{v}$ shorter than a current basis vector $\bm{b}_j$, then $\bm{b}_j$ is replaced by $\bm{v}$, provided the lattice remains unchanged. Following the original study, this process is repeated for a fixed number $\text{M}$ of times.

\begin{algorithm}[H]

  \caption{IQOAP Framework}

 ~\label{alg:IQOAP}

  \begin{algorithmic}[1]

    \Statex \textbf{Input:} Basis $B=[\bm{b}_1, \dots, \bm{b}_n]$; initial counter $iter=0$.
    \Statex \textbf{Output:} The shortest vector in the final basis $B$.
    
    \While{$iter < \text{M}$}
    
    \State Solve SVP with QAOA ansatz, obtaining the result $\bm{v}$.

    \State If $\exists j\text{ s.t. }\|\bm{v}\|<\|\bm{b}_j\|$, replace $\bm{b}_j$ with $\bm{v}$ while ensuring
    \Statex the lattice remains unchanged.

    \State If there are multiple vectors that can be replaced in
    \Statex Step 3, choose the longest one.

    \State $iter=iter+1$.
    \EndWhile

  \end{algorithmic}

\end{algorithm}

\section{~\label{sec:methodD}Iterative Partition Search Algorithm}

In this section, we present the Iterative Partition Search Algorithm (IPSA).
We will detail its two core components: the ``1-tailed search spaces" and the dynamic, stack-managed iterative process, which are designed to systematically address the limitations of previous methods. In addition, we will discuss the parameterized quantum circuit selected for its implementation on superconducting quantum processor.

\subsection{1-Tailed Search Spaces}

The 1-tailed search spaces $Y_i$ defined for an $n$-dimensional lattice as:
\begin{eqnarray}
~\label{eq:yi}
Y_i&=&\{(y_1,\cdots,y_{i-1},1,0,\cdots,0)^T\in\mathbb{Z}^n\}
\end{eqnarray}
Unlike PSA's partitions $X_i$ in Eq.~\ref{eq:xi}, the 1-tailed spaces $Y_i$ fix the $i$-th coefficient to 1. When solving SVP within partition $Y_i$, the resulting solution vector is 
\begin{eqnarray}
~\label{eq:v}
\bm{v} = \sum_{j=1}^{i-1} y_j \bm{b}_j + \bm{b}_i
\end{eqnarray}

This design, which fixes the coefficient of the $i$-th basis vector to 1, offers several crucial advantages. It reduces qubit requirements by eliminating the encoding for one coefficient. It also simplifies the cost function landscape, mitigating interference from other near-optimal vectors of very similar length.
Most critically, since the $i$-th coefficient is fixed to 1, any solution $\bm{v}$ found in $Y_i$ can directly replace $\bm{b}_i$ if $\|\bm{v}\| < \|\bm{b}_i\|$ while preserving the lattice structure, as formalized in Theorem~\ref{thm:1}. 
The ``1-tailed" strategy provides a crucial guarantee that the original IQOAP lacks: while IQOAP may successfully find a shorter vector, it cannot ensure the vector is valid for a basis update. This constrained partitioning, however, guarantees that any shorter solution found by the VQA is always valid for replacement, thereby fundamentally eliminating the problem of ineffective iterations.

\newtheorem{theorem}{Theorem}
\begin{theorem}
~\label{thm:1}
Let $B = [\bm{b}_1, \dots, \bm{b}_n]$ be a basis for a lattice $\mathcal{L}$. If a vector $\bm{v} \in \mathcal{L}$ can be expressed as $\bm{v} = \sum_{j=1}^{n} c_j \bm{b}_j$ with $c_j \in \mathbb{Z}$, and for some $k \in [1,n]$, $|c_k|=1$, then the set of vectors $B' = [\bm{b}_1, \dots, \bm{b}_{k-1}, \bm{v}, \bm{b}_{k+1}, \dots, \bm{b}_n]$ also forms a basis for $\mathcal{L}$.
\end{theorem}
\textit{Proof.}
Since $|c_k|=1$, we can express $\bm{b}_k$ as $\bm{b}_k = c_k^{-1}(\bm{v} - \sum_{j \ne k} c_j \bm{b}_j)$. As $c_k^{-1} = \pm 1$, $\bm{b}_k$ is an integer linear combination of the vectors in $B'$. Thus, $\mathcal{L}(B) \subseteq \mathcal{L}(B')$. Since $\bm{v} \in \mathcal{L}(B)$, all vectors in $B'$ are in $\mathcal{L}(B)$, so $\mathcal{L}(B') \subseteq \mathcal{L}(B)$. Therefore, $\mathcal{L}(B') = \mathcal{L}(B)$, and $B'$ is a basis for $\mathcal{L}$. $\square$

\quad

\subsection{Stack-Managed Iterative Process}

IPSA operates on an iterative principle, using a stack to manage the ``1-tailed" search partitions dynamically. This stack-managed strategy is designed to prioritize re-solving the SVP within smaller partitions immediately after a basis vector is updated.
This strategy try to ensure the basis is well-reduced (i.e., composed of shorter, more orthogonal vectors) before proceeding to larger partitions, thereby aiming to reduce the overall computational cost.
The complete algorithm is outlined in Algorithm~\ref{alg:IPSA}.

This dynamic, stack-managed strategy stands in contrast to PSA's static approach. As previously discussed, the fixed PSA's optimization landscape means that any convergence to a low-quality, suboptimal local minimum solution is worthless. However, IPSA's dynamic stack transforms these solutions into meaningful steps. As shown in Algorithm~\ref{alg:IPSA} (Lines 6-8), a successful basis update causes the algorithm to re-search for the smallest partition with the updated basis. This part of IPSA differs from PSA, which performs only a single, fixed optimization.

\begin{algorithm}[H]
\caption{Iterative Partition Search Algorithm}
~\label{alg:IPSA}
\begin{algorithmic}[1]
    \Statex \textbf{Input:} Basis $B=[\bm{b}_1, \dots, \bm{b}_n]$; initial empty stack $S$.
    \Statex \textbf{Output:} $\bm{b}_1$.
    \State Sort $B$ by increasing vector norms.
    \State Push partitions $Y_n,\cdots,Y_1$ onto $S$ ($Y_1$ at top).
    \While{$S$ is not empty}
        \State Pop partition $Y_i$ from $S$.
        \State Solve SVP for $Y_i$ using HEA, obtaining the result $\bm{v}$.
        \If{$\|\bm{v}\|<\|\bm{b}_i\|$}
        \State Replace $\bm{b}_i$ with $\bm{v}$ and keep $B$ ordered.
        \State Push partitions $Y_i,\cdots,Y_r$ onto $S$ in sequence,
        \Statex where $r$ is the position of $\bm{v}$ in $B$.
        \EndIf
        \EndWhile
\end{algorithmic}
\end{algorithm}

Since the primary focus of our work is not on deriving new optimal qubit encoding schemes. We follow the strategy demonstrated in the IQOAP framework~\cite{zhu2022iterative}, which employs a small, fixed bit allocation.

\subsection{Parameterized Quantum Circuit Selected}
\label{sec:PQC}

The design of the PQC also plays a role in VQA effectiveness. Standard ansatz like QAOA, while effective for certain optimization problems~\cite{ni2024multilevel,wu2025resource,ni2025progressive}, typically require deep circuit architectures for SVP due to the all-to-all connectivity inherent in the problem Hamiltonian (e.g., Eq.~\ref{eq:Hprime}). This limitation motivates the adoption of alternative circuit structures, such as the Hardware-Efficient Ansatz (HEA)~\cite{cong2019quantum,henderson2020quanvolutional,liu2021hybrid,hur2022quantum,herrmann2022realizing}, which can provide strong expressive power with shallower circuits. Recently, we have also noted other mitigation strategies, such as using Fixed-Angle QAOA variants~\cite{boulebnane2024solving,prokop2025heuristic,priestley2025practically} or leveraging classical post-processing~\cite{hou2025effcient}.

\begin{figure}[!htb]
\centering
\includegraphics[width=\columnwidth]{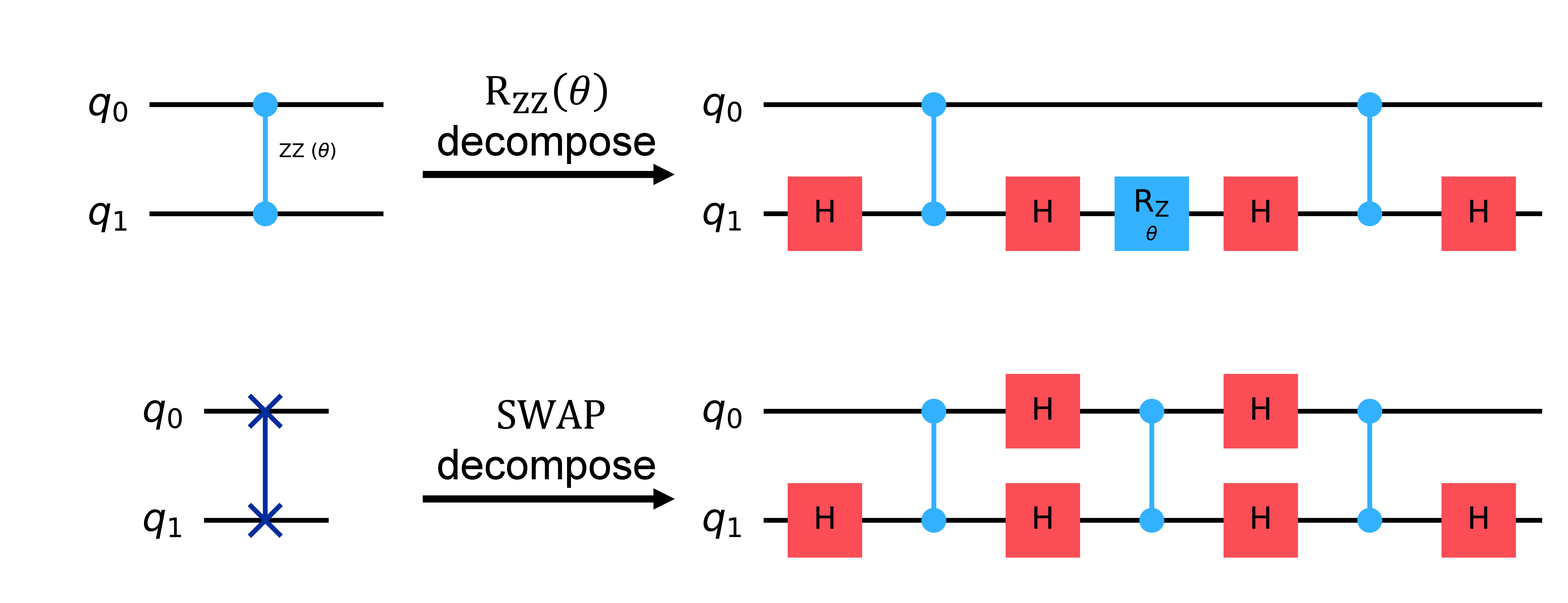}
\caption{~\label{Fig:decompose}\raggedright
Decomposition of logical gates required by the QAOA ansatz. An $\text{R}_{\text{ZZ}}(\theta)$ rotation gate is decomposed into a sequence involving two CZ gates. A SWAP gate is decomposed into three CZ gates.}
\end{figure}

On the superconducting quantum processor, this all-to-all connectivity requirement presents a significant challenge during transpilation. The QAOA ansatz, when applied to the problem Hamiltonian, must implement entangling operations such as $\text{R}_{\text{ZZ}}(\theta)$ gates between all pairs of qubits. These logical entangling gates typically decompose into CZ gates. However, current superconducting quantum processors have limited qubit connectivity, with all-to-all connections not physically available. To execute gates between non-adjacent qubits, the transpiler must insert a substantial number of SWAP gates, which themselves also decompose into multiple CZ gates (Fig.~\ref{Fig:decompose}). This process results in a significant ``transpile tax'' overhead. The resulting physical circuit depth increases significantly, making the algorithm highly susceptible to gate errors and decoherence within the device's finite coherence times.

\begin{figure*}[!htb]
\centering
\includegraphics[width=0.95\textwidth]{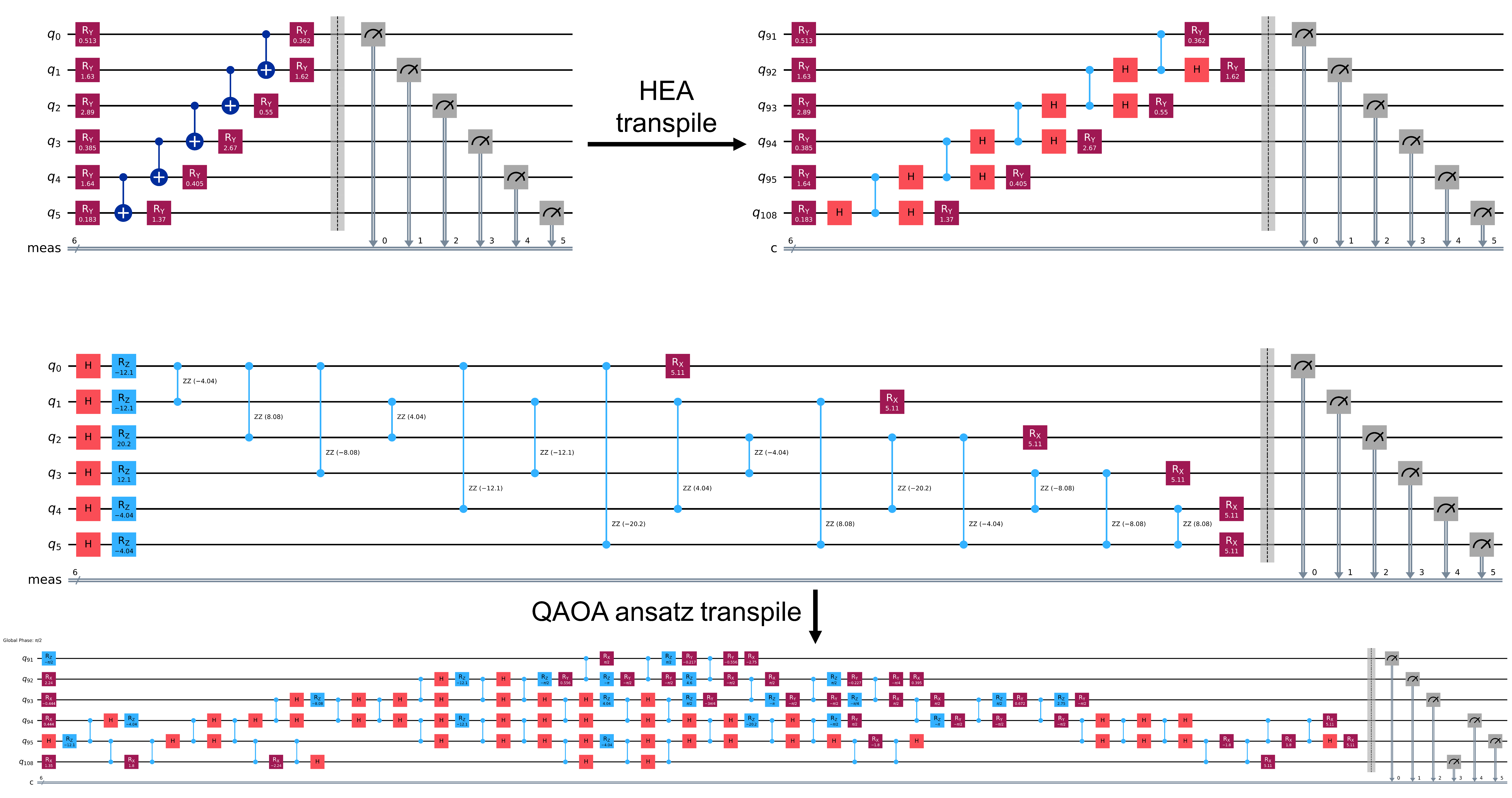}
\caption{~\label{Fig:transpile}\raggedright
Visualization of the compilation overhead for different ansatz structures. The logical circuit diagrams are shown alongside their corresponding physical circuits after transpilation for a linear connectivity topology. The all-to-all connectivity requirement of the QAOA ansatz, which is optimized by Qiskit's transpiler, still results in a deep physical circuit. The HEA, designed to match the linear topology, undergoes minimal change during transpilation.}
\end{figure*}

In contrast, the HEA structure is Hamiltonian-agnostic. Its entangling blocks can be designed to align with the native hardware topology. This alignment drastically reduces the required number of SWAP gates during transpilation. Consequently, the physical HEA circuit remains comparatively shallow and is more robust to hardware noise. This stark contrast in compilation outcomes is visualized in Fig.~\ref{Fig:transpile}, and the quantitative resource comparison is detailed in Table~\ref{tab:table1}. We provide a detailed performance comparison between QAOA and HEA under ideal, noiseless simulation conditions in Appendix~\ref{app:A}, which further supports our selection. Therefore, we utilize the HEA as the PQC structure for IPSA.

\begin{table}[b]

\caption{~\label{tab:table1}\raggedright
A comparison of the logical and transpiled 1-layer QAOA ansatz and the 1-layer HEA for a 6-qubit implementation to be run on real hardware in terms of depth and number of entangling gates. We consider both unoptimized and optimized transpilation using Qiskit’s transpiler (optimization levels 0 and 3) before execution on the \textit{Baihua} superconducting quantum processor via the Quafu platform.}

\begin{ruledtabular}
\begin{tabular}{l c c}
\multicolumn{3}{c}{\textrm{Logical Circuits}} \\
\textrm{Ansatz} & \textrm{Depth} & \textrm{Entangling Gates} \\
\colrule
QAOA & 12 & 15 \\
HEA  & 7 & 5 \\
\hline

\multicolumn{3}{c}{\textrm{Transpiled Circuits}} \\
\multicolumn{3}{c}{\textrm{(Basis Gate: H, Rx, Ry, Rz, CZ)}} \\
\textrm{Ansatz} & \textrm{Depth} & \textrm{CZ Gates} \\
\colrule
QAOA (unoptimized) & 92 & 51 \\
QAOA (optimized)   & 64 & 43 \\
HEA (unoptimized)  & 13 & 5 \\
HEA (optimized)    & 13 & 5 \\
\end{tabular}
\end{ruledtabular}

\end{table}

\section{Real Hardware Experiments}
\label{sec:hardware_exp}
In this section, we present the experimental validation of our proposed IPSA. We benchmark its performance against IQOAP and PSA, through execution on the \textit{Baihua} superconducting quantum processor, which is accessed via the Quafu platform~\cite{Quafu}. Detailed specifications of this hardware and its topology are provided in Appendix~\ref{app:B}. We will first detail the complete experimental setup and subsequently present the comparative results obtained under realistic noise conditions.

\subsection{Experimental Setup}
\label{sec:setup}
For the experimental validation, we generated a test set of 50 random SVP instances in dimension $n=4$. This dimension was selected as a representative case for hardware execution, given the significant impact of device noise, limited coherence times, and overall execution time constraints on larger-scale problems.
These instances were created using the standard procedure of applying random unimodular transformations to LLL-reduced lattice bases~\cite{zhu2022iterative,barbera2024finding}.

For the qubit allocation, reflecting the qubit-saving characteristic of iterative frameworks, both IPSA and IQOAP encode each variable coefficient using two qubits, whereas PSA utilizes three qubits per coefficient. For the $n=4$ instances, this results in a circuit with 6 qubits for IPSA, 8 for IQOAP, and 11 for PSA. To ensure a fair comparison, the coefficient range for all test instances was restricted to be representable within the 3-qubit allocation used by PSA.

Regarding the PQC structure, both IPSA and PSA employ a 1-layer HEA. The IQOAP implementation uses a 1-layer QAOA with parameters constrained such that $\beta=\gamma$, consistent with its original proposal~\cite{zhu2022iterative}.

\begin{figure*}[!htb]
\centering
\includegraphics[width=\textwidth]{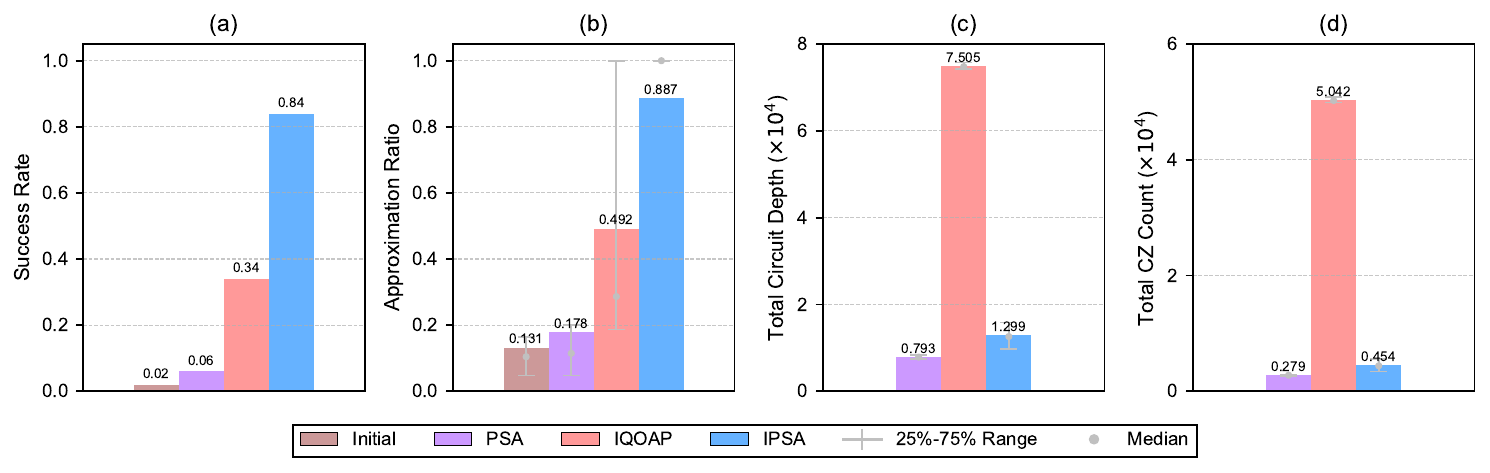}
\caption{~\raggedright\label{fig:hardware_results}Comparative analysis of IPSA (blue), PSA (purple), and IQOAP (red) on the 50 SVP instances at $n=4$, executed on the \textit{Baihua} superconducting quantum processor. The panels show (a) Success Rate (SR), (b) Approximation Ratio (AR), (c) Total Circuit Depth ($D_{\text{total}}$), and (d) Total CZ Count ($C_{\text{total}}$). For the data points, the error bars indicate the interquartile range (25th to 75th percentile), and the markers denote the median values.}
\end{figure*}

To quantitatively assess algorithm performance, we utilize the following metrics:
\begin{enumerate}

    \item \textbf{Success Rate (SR):} The fraction of instances where the algorithm successfully finds the shortest vector, defined as $SR = N_{\text{succ}} / N_{\text{total}}$.

    \item \textbf{Approximation Ratio (AR) and Average Approximation Ratio (AAR):} The quality of the found solution for a given instance, defined as $AR = \lambda_1(\mathcal{L}) / \|v_{\text{alg}}\|$, where $\lambda_1(\mathcal{L})$ is the length of the shortest vector and $\|v_{\text{alg}}\|$ is the length of the vector found by the algorithm. An AR value ranges from 0 to 1, with a value closer to 1 indicating a higher-quality solution. The AAR is then calculated for each algorithm and dimension by taking the mean of the AR values over all instances.

    \item \textbf{Total Circuit Depth ($D_{\text{total}}$):} The cumulative circuit depth across all iterations of the algorithm, $D_{\text{total}} = \sum_{i=1}^{I} d_i$, where $d_i$ is the depth of the circuit in iteration $i$. This metric reflects, to some extent, the total execution time on quantum hardware.

    \item \textbf{Total CZ Count ($C_{\text{total}}$):} The cumulative number of two-qubit CZ gates used across all iterations, $C_{\text{total}} = \sum_{i=1}^{I} c_i$. This metric quantifies the primary source of noise and error in many NISQ devices.

\end{enumerate}

\subsection{Experimental Results and Analysis}

We now present the results from the experimental execution on the \textit{Baihua} superconducting quantum processor, comparing IPSA against PSA and IQOAP. The performance of the three algorithms on the 50 test instances is summarized in Fig.~\ref{fig:hardware_results}.

As shown in Fig.~\ref{fig:hardware_results}(a), IPSA achieves a Success Rate (SR) of 0.84, which is approximately 2.5 times the SR of 0.34 obtained by IQOAP. In contrast, the SR for PSA was significantly lower at 0.06. Regarding solution quality, IPSA also demonstrates strong performance, reaching an Average Approximation Ratio (AAR) of 0.887. This value is roughly 1.8 times higher than the 0.492 achieved by IQOAP and 5 times higher than the 0.178 of PSA, as depicted in Fig.~\ref{fig:hardware_results}(b).

The hardware experiments also highlight the resource efficiency of IPSA. As detailed in Fig.~\ref{fig:hardware_results}(c), the $D_{\text{total}}$ for IPSA was $1.299 \times 10^4$, representing an 82.7\% reduction compared to the $7.505 \times 10^4$ required by IQOAP. This significant difference is a direct consequence of the heavy ``transpile tax'' on real hardware, as discussed in Sec.~\ref{sec:PQC}. A similar trend is observed for $C_{\text{total}}$ in Fig.~\ref{fig:hardware_results}(d), where IPSA uses 91.0\% fewer CZ gates compared to the $5.042 \times 10^4$ count of IQOAP.

In summary, the experimental results on real quantum hardware demonstrate that IPSA achieves a superior balance between solution quality and resource consumption on current superconducting quantum processors. Compared to PSA, IPSA achieves a 14-fold increase in SR with less than double the total circuit depth. Conversely, compared to IQOAP, IPSA reduces $D_{\text{total}}$ by approximately 83\% while achieving approximately 2.5 times the SR of IQOAP. Since $D_{\text{total}}$ reflects the cumulative quantum circuit execution time to some extent, this reduction suggests a significant advantage for IPSA in overall runtime efficiency when compared to IQOAP.
Simultaneously, the high SR relative to PSA indicates that the dynamic, stack-managed IPSA is more resilient to local minima than static search methods, even under realistic noise conditions. These outcomes suggest that by addressing the structural inefficiencies of its predecessors, IPSA provides a more practical path for applying iterative VQAs to combinatorial optimization problems on near-term quantum devices.

In addition, these real hardware results are consistent with findings from extensive noiseless numerical simulations presented in Appendix~\ref{app:C}, which demonstrate similar performance advantages for IPSA on larger and more challenging sets of SVP instances.

\qquad

\section{~\label{sec:con}Conclusion}
In this work, we proposed IPSA, a ``second-generation" iterative VQA-based SVP solver 
that advances the frameworks of IQOAP and PSA.
It is designed to overcome their fundamental drawbacks, including ineffective iterations, oversized search spaces, and low circuit efficiency. Its core features, the 1-tailed search spaces and a dynamic, stack-managed iterative process, are the key components for achieving this goal. Although PSA and IQOAP may not be the best approach for solving SVP with VQA today, 
by systematically addressing these known limitations, the refined IPSA presented in this work emerges as a competitive approach.
Furthermore, IPSA can be employed as a subroutine within algorithms like as block Korkin-Zolotarev or applied to problems such as learning with errors, providing an efficient and reliable core component for these complex tasks.

\begin{acknowledgments}
This work is supported by National Natural Science Foundation of China (Grant Nos. U25B2014, 62372048, 62272056, 62371069), and National Key Laboratory of Security Communication Foundation (2025, 6142103042503). We acknowledge the support of Quafu Quantum Cloud Computing Platform from Beijing Academy of Quantum Information Sciences (https://quafu-sqc.baqis.ac.cn/).
\end{acknowledgments}
\bibliography{IPSA_revised}
\begin{widetext}

\appendix

\section{Comparative Analysis of HEA and QAOA Ansatz under Ideal Simulation}
\label{app:A}

This appendix provides an empirical validation for the selection of the HEA over the QAOA ansatz. The comparisons presented here are conducted using ideal, noiseless numerical simulations which assume all-to-all qubit connectivity. The definitions for the performance metrics used in this section are provided in Sec.~\ref{sec:setup}.

Acknowledging the different expressive power of the two structures, we compare the relative effectiveness of the IPSA implementation using a 
1-layer
HEA against the version using a 4-layer QAOA. The comparison was conducted on a set of 600 SVP instances (200 for each dimension $n \in \{4, 5, 6\}$). These instances were created using a standard procedure: applying random unimodular transformations to the reduced lattice basis~\cite{zhu2022iterative,barbera2024finding}. The results are presented in Fig.~\ref{fig:app_qaoahea}

\begin{figure*}[!htb]
\centering
\includegraphics[width=\textwidth]{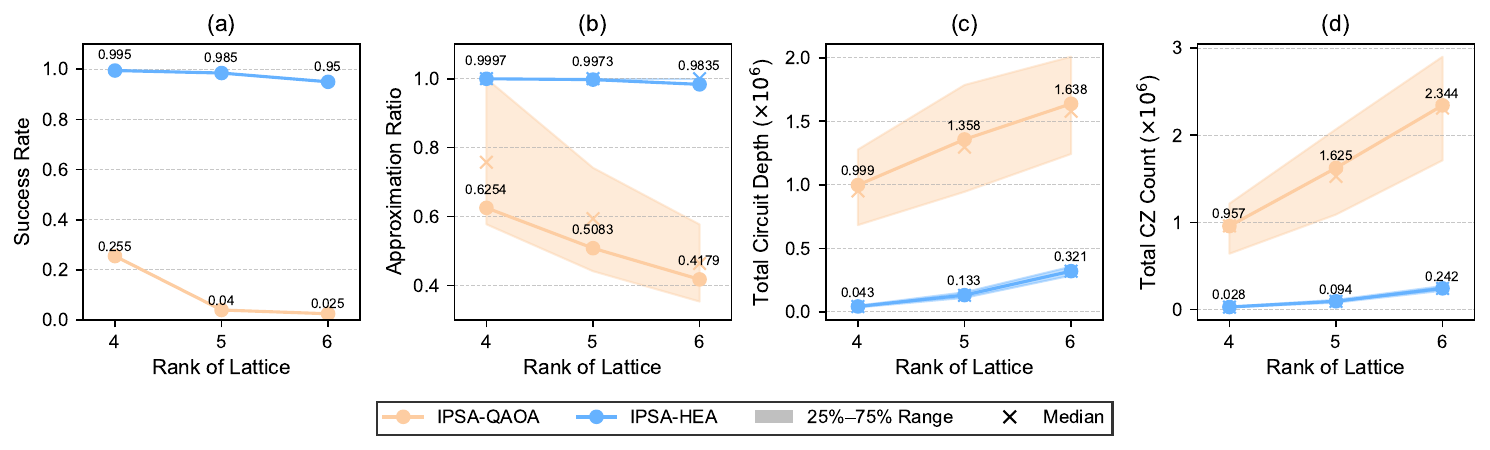}
\caption{~\label{fig:app_qaoahea}\raggedright Comparative analysis of IPSA-HEA (blue) and IPSA-QAOA (orange) on the set of 600 SVP instances for dimensions $n \in \{4, 5, 6\}$. The four panels show (a) Success Rate (SR), (b) Approximation Ratio (AR), (c) Total Circuit Depth ($D_{\text{total}}$), and (d) Total CZ Count ($C_{\text{total}}$). The y-axis values for panels (c) and (d) are presented in units of $10^6$. In all plots, the shaded areas represent the interquartile range (25th to 75th percentiles), and the cross markers indicate the median values.}
\end{figure*}

The figure shows a clear performance difference when using the HEA implementation across all dimensions. For instance, at $n=6$, the HEA variant achieved an SR of 0.95, whereas the QAOA variant's SR was only 0.025. The AAR for IPSA-HEA also remained high and close to 1. Its AAR value for $n=6$ was 0.9835. In contrast, the AAR for IPSA-QAOA decreased as the problem size increased, falling from 0.6254 at $n=4$ to 0.4179 at $n=6$.

Furthermore, the HEA-based approach was substantially more resource-efficient in simulation. At $n=6$, $D_{\text{total}}$ for IPSA-HEA was $0.321\times 10^6$. This result is approximately five times lower than the $1.638\times 10^6$ required by IPSA-QAOA. Its $C_{\text{total}}$ showed an even greater difference. The value was $0.242\times 10^6$ for IPSA-HEA, almost ten times lower than the $2.344\times 10^6$ for the QAOA variant. Similar resource advantages for the HEA implementation were observed for $n=4$ and $n=5$.

We note that while the HEA variant's total cost is substantially lower in the tested range, its scaling trend appears steeper than that of the QAOA variant. This can be viewed as the necessary computational trade-off for an ansatz capable of achieving a high success rate, whereas the 4-layer QAOA variant fails to find a solution. We acknowledge that a detailed analysis of the precise asymptotic scaling for both ansatzes is a subject for future investigation.

In summary, these simulation results provide strong empirical evidence that for the SVP instances under consideration within the IPSA, the HEA offers a clear advantage over the standard QAOA ansatz. It delivers superior solution quality with significantly lower resource consumption under ideal conditions. This results complements the hardware-compilation analysis in the main text and supports the adoption of HEA as the PQC for IPSA.

\section{Hardware Specifications}
\label{app:B}

In this appendix, we provide additional details about the experimental hardware used in the main text. The experiments were executed on the \textit{Baihua} superconducting quantum processor, accessible via the Quafu Superconducting Quantum Computing cloud~\cite{Quafu}. This processor features 156 superconducting qubits. The qubit connectivity map for two-qubit entangling gates is shown in Fig.~\ref{fig:baihua_topo}. The median calibration data for the device, retrieved at the time of the experiments, is summarized in Table~\ref{tab:baihua_calib}.

\begin{figure}[!htb]
\centering
\includegraphics[width=\columnwidth]{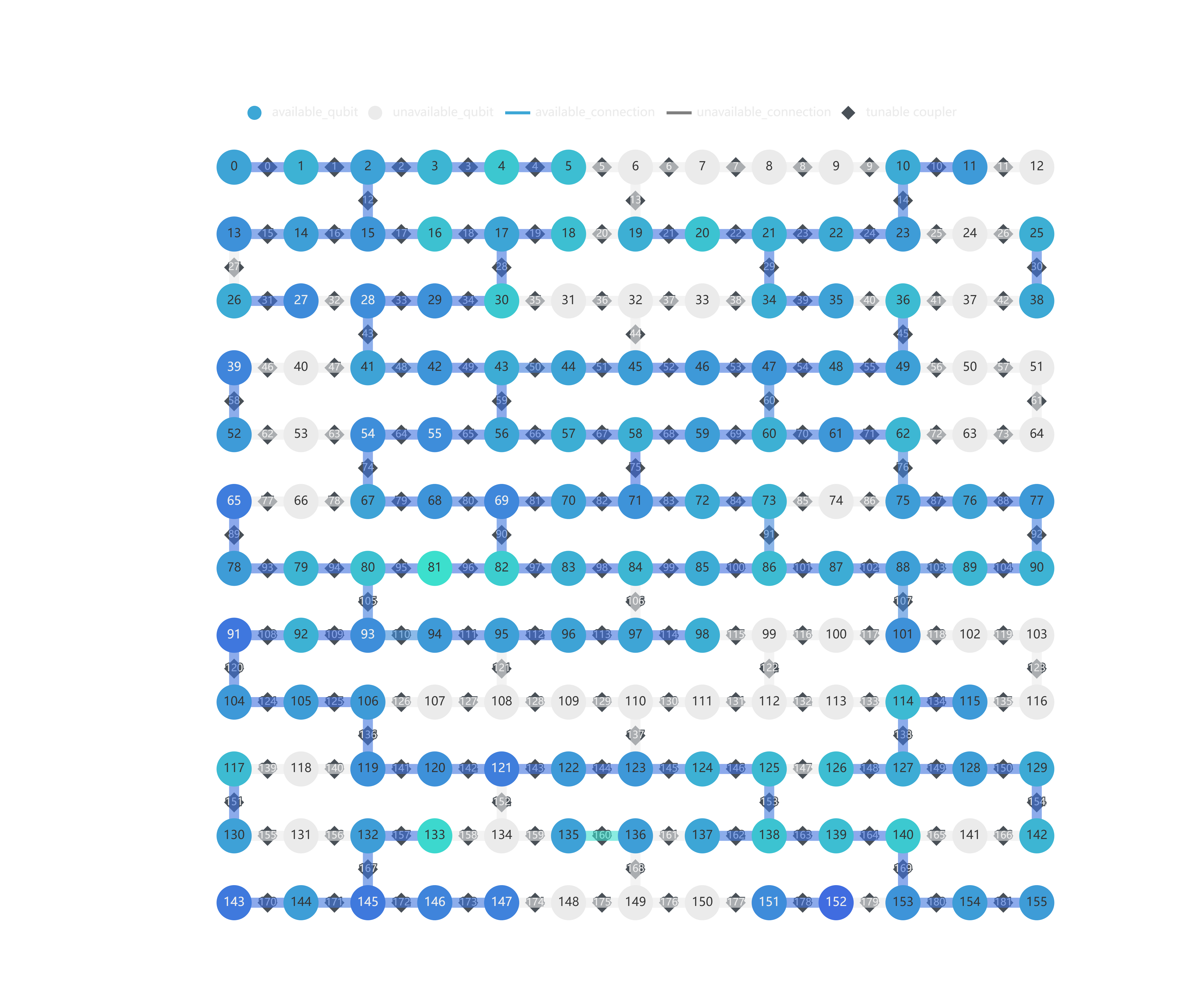}
\caption{~\label{fig:baihua_topo}\raggedright
The qubit connectivity topology of the \textit{Baihua} superconducting quantum processor. Blue circles indicate available qubits, while light gray circles represent unavailable qubits. Blue arrows denote available connections for two-qubit gates.}
\end{figure}

\begin{table}[h]
\caption{\raggedright
Median calibration data for the \textit{Baihua} superconducting quantum processor, representative of the device performance during the experimental period.}
\label{tab:baihua_calib}
\centering
\begin{minipage}{0.75\linewidth}
\begin{ruledtabular}
\begin{tabular}{l c}
\textrm{Parameter} & \textrm{Value} \\
\colrule
Experimental period & October 23 -- December 22, 2025 \\
Basis gates & H, Rx, Ry, Rz, CZ \\
Median T1 & 71.608 $\mu$s \\
Median T2 & 23.724 $\mu$s \\
Median 1Q error & 7.6e-4 \\
Median 2Q error & 1.4e-2 \\
\end{tabular}
\end{ruledtabular}
\end{minipage}
\end{table}

\section{Extended Numerical Simulations}
\label{app:C}
In this appendix, we present comprehensive numerical simulations that extend the experimental validation provided in the main text. Specifically, we compare IPSA against IQOAP and multiple variants of PSA (3-, 4-, and 5-qubit encodings) on a larger-scale Benchmark Set. Furthermore, we evaluate algorithm performance on a specialized ``LLL-Challenging Set" designed to test robustness against strong suboptimal solutions.

\subsection{Instances Generation and Setup}
\label{app:C1}
The noiseless numerical simulations utilize two distinct sets of SVP instances, totaling 800 instances.
\begin{itemize}
    \item \textbf{Benchmark Set:} Comprises 600 instances (200 each for dimension $n \in \{4, 5, 6\}$). These were generated by applying random unimodular transformations to reduced lattice bases. Crucially, to ensure a rigorous evaluation of search capabilities, we employed unimodular matrices with larger entries than main-text instances, thereby creating more challenging lattice bases with highly skewed geometries.
    \item \textbf{LLL-Challenging Set:} Consists of 200 instances at $n=6$. These instances were specifically engineered by selecting random lattices where the classical LLL algorithm fails to find the true shortest vector, instead converging to a suboptimal vector that is, on average, only $\approx 2\%$ longer. This set critically tests an algorithm's ability to discriminate between the global optimum and close local minima.
\end{itemize}

We expanded the comparative analysis to include variants of the PSA algorithm with different qubit allocations, denoted as $k$-PSA, where each coefficient is encoded using $k \in \{3, 4, 5\}$ qubits. Due to the prohibitive classical simulation costs for circuits exceeding 22 qubits, the 5-PSA variant was excluded for $n=5$, and both 4-PSA and 5-PSA were excluded for $n=6$. The key parameters for all simulated algorithms are summarized in Table~\ref{tab:sim_params}.

\begin{table}[h]
\caption{\raggedright\label{tab:sim_params}Parameter configurations for the algorithms evaluated in the noiseless numerical simulations. The settings for IQOAP are consistent with its original study~\cite{zhu2022iterative}.}
\centering
\begin{minipage}{0.75\linewidth}
\begin{ruledtabular}
\begin{tabular}{cccc}
\textrm{Algorithms}&
\textrm{SVP dimension}&
\textrm{Max. Qubits}&
\textrm{PQC}\\
\colrule
\multirow{3}{*}{IPSA} & 4 & 6 & \multirow{3}{*}{HEA} \\
                      & 5 & 8 &                   \\
                      & 6 & 10 &                  \\
\colrule
\multirow{3}{*}{IQOAP} & 4 & 8 & \multirow{3}{*}{QAOA($\beta=\gamma$)} \\
                       & 5 & 10 &                                    \\
                       & 6 & 12 &                                    \\
\colrule
\multirow{3}{*}{3-PSA} & 4 & 11 & \multirow{3}{*}{HEA} \\
                       & 5 & 14 &                   \\
                       & 6 & 17 &                   \\
\colrule
\multirow{2}{*}{4-PSA} & 4 & 15 & \multirow{2}{*}{HEA} \\
                       & 5 & 19 &                   \\
\colrule
\multirow{1}{*}{5-PSA} & 4 & 19 & \multirow{1}{*}{HEA} \\
\end{tabular}
\end{ruledtabular}
\end{minipage}
\end{table}

For the classical optimization component of all algorithms, we employed the \texttt{minimize} function from the \texttt{SciPy.optimize} library, selecting Powell's conjugate direction method as the optimizer. The initial parameters for the rotational gates in the PQC were randomly initialized from a uniform distribution over the interval $[0, \pi]$. We adopted the default termination criteria from the SciPy implementation: the optimization process stops when the fractional tolerance in either the parameters (\texttt{xtol}) or the cost function value (\texttt{ftol}) is below $1 \times 10^{-4}$. All simulations were performed on the Qiskit framework using the \texttt{StatevectorEstimator} or \texttt{StatevectorSampler}. The definitions for the performance metrics used in this section are provided in Sec.~\ref{sec:setup}, as same as the main text uses.

\subsection{Results on the Benchmark Set}
\label{app:C2}
We now compare the effectiveness of IPSA against the various $k$-PSA and IQOAP. The complete results are summarized in Fig.~\ref{Fig:result}.

\begin{figure}[!htb]
\centering
\includegraphics[width=\textwidth]{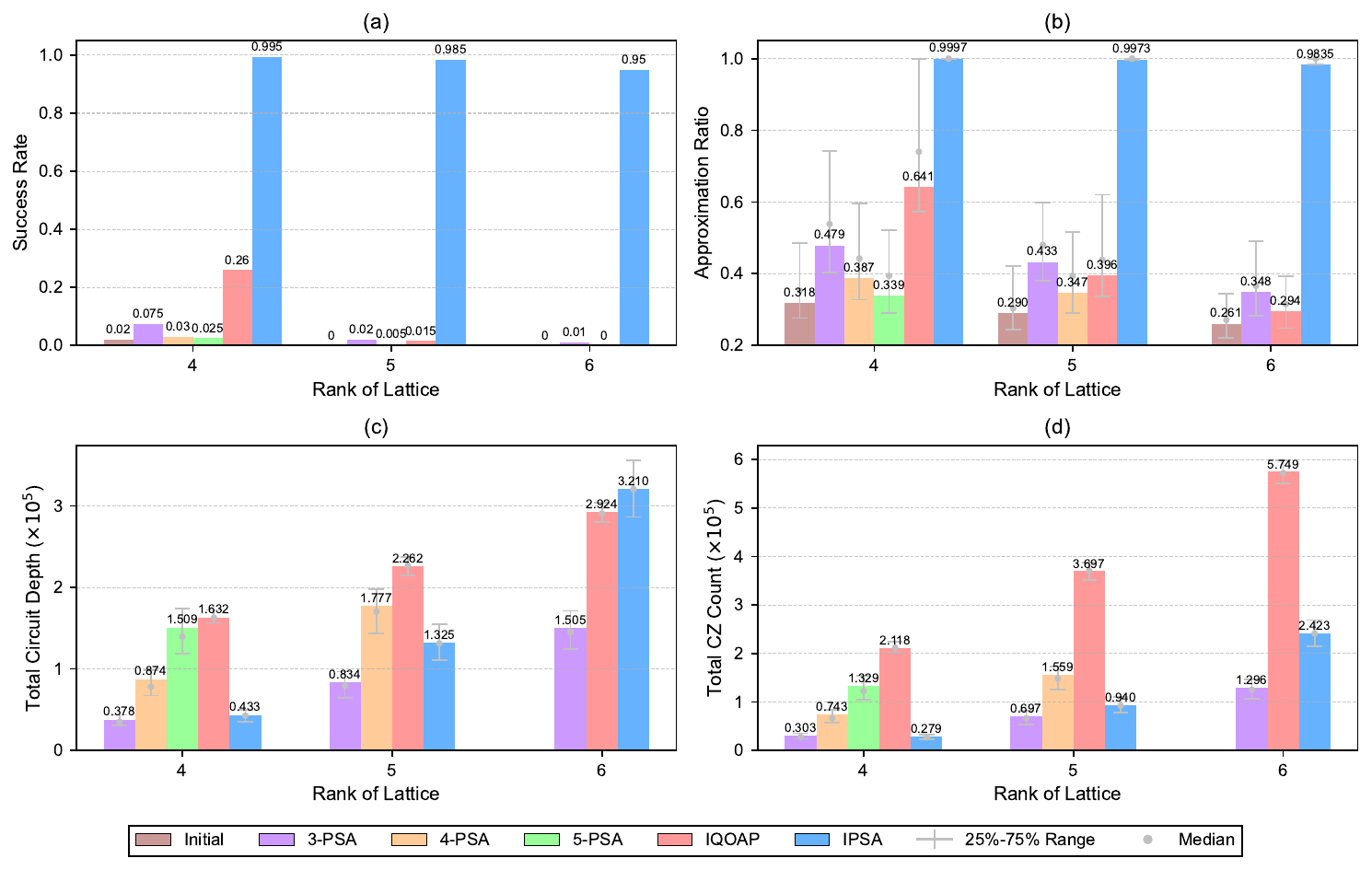}
\caption{~\raggedright\label{Fig:result}Comparative analysis of IPSA (blue) and several existing algorithms, including the $k$-PSA variants (3-PSA in purple, 4-PSA in orange, and 5-PSA in green) and IQOAP (red). The comparison is across dimensions $n \in \{4, 5, 6\}$ on the Benchmark Set. The panels show (a) Success Rate (SR), (b) Approximation Ratio (AR), (c) Total Circuit Depth ($D_{\text{total}}$), and (d) Total CZ Count ($C_{\text{total}}$). The y-axis values for (c) and (d) are in units of $10^5$. In panels (a) and (b), the brown bars labeled ``Initial" represent the metrics of the initial basis. For the data points, the error bars indicate the interquartile range (25th to 75th percentile), and the circular markers denote the median values.}
\end{figure}

Regarding the primary goal of finding the shortest vector, IPSA substantially outperforms the other algorithms. As shown in Fig.~\ref{Fig:result}(a), for dimensions $n=4$, 5, and 6, IPSA achieved an SR of 0.995, 0.985, and 0.95, respectively. By comparison, the SR for IQOAP showed a significant decline from 0.26 to 0.01 across the exact dimensions. The $k$-PSA variants yielded comparatively lower success rates, with SR values generally below 0.1.

We note that the SR for IQOAP in these noiseless simulations is lower than that observed in the real hardware experiments presented in the main text. This discrepancy arises from the increased difficulty of the instances used in this appendix. As detailed in Appendix~\ref{app:C1}, these simulations employed unimodular matrices with larger entries to create more complex lattice geometries, whereas the hardware experiments in the main text used instances with a restricted coefficient range. Even in this noiseless environment, the performance of IQOAP and $k$-PSA degrades significantly on these more challenging instances, whereas IPSA maintains a high SR. This outcome suggests that IQOAP's fixed iteration limit is insufficient for these more complex instances. Although increasing the number of iterations might marginally improve the success rate, this reliance on a fixed iteration count appears to be a structural limitation that hinders scalability, as evidenced by the rapid performance decline at $n=5$ and $n=6$.


The resource requirements present a more varied picture. For $n=4$ and 5, the $D_{\text{total}}$ of IPSA was comparable to or lower than that of IQOAP and the higher-qubit $k$-PSA variants. For the largest dimension $n=6$, the $D_{\text{total}}$ for IPSA at $0.321\times10^6$ was higher than for 3-PSA and IQOAP. IQOAP's lower depth, however, is mainly attributable to its termination after a fixed 50 iterations, a number insufficient to reliably solve the problem at a little larger scale. In terms of the $C_{\text{total}}$, IPSA consistently required fewer CZ gates than IQOAP but more than the PSA variants. This increased resource consumption is a trade-off for the substantial gains in success rate and solution quality.

Crucially, it is important to note that the $D_{\text{total}}$ of IQOAP is comparable to that of IPSA in these numerical simulations because the results reflect only the logical circuit depth. This metric excludes the heavy ``transpile tax'' incurred by limited connectivity on real quantum computers, which explains the depth parity observed in this ideal simulation environment.

The results indicate that both PSA and IQOAP find the shortest vector with low probability, making them unreliable for this task. This low reliability stems from PSA's tendency to converge on local minima and IQOAP's propensity for ineffective iterative refinements that fail to improve solution quality. Their limited success at a modest dimension of $n=6$ suggests poor scalability for higher-dimensional problems. In contrast, IPSA achieves the correct solution with high probability. Its integration of iterative refinement and a novel partitioning strategy demonstrates a more effective path to achieving high-quality solutions for the SVP.

\subsection{Results on the LLL-Challenging Set}
\label{app:C3}
We then assessed IPSA's effectiveness on the LLL-Challenging Set. The goal was to evaluate its ability to distinguish the shortest vector from close suboptimal solutions, which is a key difficulty in solving the SVP. These 200 6-dimensional instances from the LLL-Challenging Set are characterized by having LLL-reduced solutions that are, on average, only about 2\% longer than the shortest vector length. We compared IPSA against 3-PSA and IQOAP, with the results shown in Fig.~\ref{Fig:LLL}.

\begin{figure}[!htb]
\centering
\includegraphics[width=\textwidth]{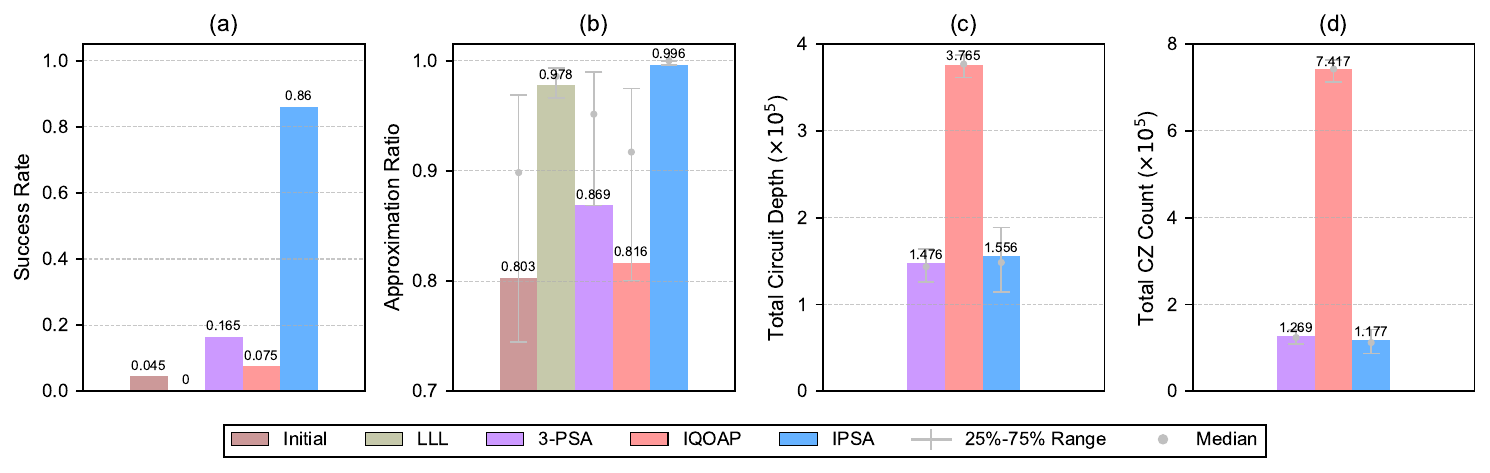}
\caption{~\raggedright\label{Fig:LLL}Comparative analysis of IPSA (blue), 3-PSA (purple), and IQOAP (red) on the 200 instances of the LLL-Challenging Set ($n=6$). The classical LLL algorithm (yellow-green) is also included for reference. The panels show (a) Success Rate (SR), (b) Approximation Ratio (AR), (c) Total Circuit Depth ($D_{\text{total}}$), and (d) Total CZ Count ($C_{\text{total}}$). The y-axis values for (c) and (d) are in units of $10^5$. By the selection criteria for this instance set, the SR for the LLL algorithm is zero, but its average AR is high at 0.978.}
\end{figure}

As illustrated in the figure, IPSA maintained a high level of solution quality even in the presence of strong suboptimal attractors. It achieved an SR of 0.86. This result was substantially higher than the SR of 0.165 for 3-PSA and 0.075 for IQOAP. In terms of solution quality, IPSA reached an AAR of 0.996. This result outperforms the AAR values from 3-PSA at 0.869, IQOAP at 0.816, and even the classical LLL algorithm's initial result of 0.978 in these instances. It is noteworthy that the AAR for 3-PSA and IQOAP improved on this set compared with the $n=6$ Benchmark Set. This phenomenon is because the primary goal of the LLL-Challenging Set was to test an algorithm's discrimination of close suboptimal solutions. Therefore the instances were generated without the randomizing unimodular transformations used for the Benchmark Set. However, the large error bars associated with both 3-PSA and IQOAP in Fig.~\ref{Fig:LLL}(b) indicate significant instability in their results across the instance set. This outcome highlights IPSA's ability to maintain high solution fidelity, even when the solution space contains a prominent local minimum corresponding to the LLL-reduced vector.

This level of solution quality did not require a significant increase in computational overhead compared with the simpler methods. The $D_{\text{total}}$ for IPSA was $1.556\times10^5$, which is comparable to the value for 3-PSA at $1.476\times10^5$. The $C_{\text{total}}$ for IPSA was $1.177\times10^5$, a value also comparable to the $1.269\times10^5$ for 3-PSA. Both of IPSA's resource metrics were considerably lower than those for IQOAP.

The ability to distinguish between vectors of very similar lengths is a critical hurdle for VQA-based SVP solvers. Heuristic methods are susceptible to being trapped in local minima corresponding to these near-shortest vectors. This issue is particularly pronounced in the LLL-Challenging Set, where the LLL solution and the actual shortest vector reside in similar search regions. IPSA's 1-tailed partitioning strategy is designed to address this challenge. By design, the strategy separates such closely matched candidate vectors into different search stages or ensures they are differentiated through the deterministic basis update and sorting mechanism of the algorithm. This transformation of a difficult heuristic search into a sequence of more defined sub-problems is central to IPSA's effectiveness. It allows the algorithm to collect multiple distinct candidate solutions in different iterations, enabling the identification of the global optimum through a final, deterministic comparison of their vector lengths.

The effectiveness of IPSA on the LLL-Challenging Set highlights the efficacy of its 1-tailed search spaces strategy. This design feature equips the algorithm to perform well in scenarios with substantial suboptimal solution interference, a valuable capability for solving the SVP and other challenging combinatorial optimization problems.

\subsection{Summary of Numerical Findings}
The extensive numerical simulations reinforce and extend the findings from the hardware experiments. They confirm that IPSA's advantages in success rate and solution quality are not limited to small-scale instances but scale to harder, higher-dimensional problems effectively. The comparison with multiple PSA variants demonstrates that IPSA's iterative, stack-managed architecture provides a fundamental algorithmic advantage over static approaches, regardless of their specific qubit resource allocation.

\end{widetext}

\end{document}